\documentclass{amsart}
\usepackage{hyperref}

\newtheorem{theorem}{Theorem}[section]
\newtheorem{lemma}[theorem]{Lemma}
\newtheorem{remark}[theorem]{Remark}

\newcommand{\R}{{\mathbb R}}
\newcommand{\N}{{\mathbb N}}
\newcommand{\Z}{{\mathbb Z}}
\newcommand{\C}{{\mathbb C}}
\newcommand{\M}{{\mathbb M}}

\newcommand{\nn}{\nonumber}
\newcommand{\be}{\begin{equation}}
\newcommand{\ee}{\end{equation}}
\newcommand{\bea}{\begin{eqnarray}}
\newcommand{\eea}{\end{eqnarray}}
\newcommand{\ul}{\underline}
\newcommand{\ol}{\overline}
\newcommand{\ti}{\tilde}

\newcommand{\I}{\mathrm{i}}

\newcommand{\tr}{\mathrm{tr}}
\newcommand{\im}{\mathrm{Im}}
\newcommand{\re}{\mathrm{Re}}

\newcommand{\vrc}{\ul{\Xi}_{p_0}}

\newcommand{\di}{\mathcal{D}}

\newcommand{\Amap}{\ul{A}_{p_0}}
\newcommand{\amap}{\ul{\alpha}_{p_0}}

\newcommand{\hamap}{\ul{\hat{\alpha}}_{p_0}}

\newcommand{\Rg}[1]{R_{2g+2}^{1/2}(#1)}

\newcommand{\eps}{\varepsilon}

\newcommand{\lam}{\lambda}

\newcommand{\om}{\omega}

\newcommand{\cprime}{\/{\mathsurround=0pt$'$}}

\numberwithin{equation}{section}

\begin{document}

\title[Algebro-Geometric Constraints on Solitons]{Algebro-Geometric Constraints on Solitons with Respect to Quasi-Periodic Backgrounds}

\author[G. Teschl]{Gerald Teschl}
\address{Faculty of Mathematics\\
Nordbergstrasse 15\\ 1090 Wien\\ Austria\\ and International Erwin Schr\"odinger
Institute for Mathematical Physics, Boltzmanngasse 9\\ 1090 Wien\\ Austria}
\email{\href{mailto:Gerald.Teschl@univie.ac.at}{Gerald.Teschl@univie.ac.at}}
\urladdr{\href{http://www.mat.univie.ac.at/~gerald/}{http://www.mat.univie.ac.at/\~{}gerald/}}

\thanks{{\it To appear in Bull. London Math. Soc.}}
\thanks{{\it Supported by Austrian Science Fund (FWF) under Grant No.\ P17762}}


\begin{abstract}
We investigate the algebraic conditions that have to be satisfied by the scattering
data of short-range perturbations of quasi-periodic finite-gap Jacobi operators
in order to allow solvability of the inverse scattering problem.
Our main result provides a Poisson-Jensen-type formula for
the transmission coefficient in terms of Abelian integrals on the underlying
hyperelliptic Riemann surface and an explicit condition for its
single-valuedness. In addition, we establish trace formulas which
relate the scattering data to the conserved quantities in this case.
\end{abstract}

\maketitle

\section{Introduction}

Solitons are a key feature of completely integrable wave equations and
there are usually two ways of constructing the $N$-soliton solution
with to respect to a given background solution. Both are based
on fact that the underlying Lax operator is reflectionless with respect
to the background, but has $N$ additional eigenvalues. One is via the inverse
scattering transform by choosing an arbitrary number of eigenvalues
(plus corresponding norming constants) and setting the reflection
coefficient equal to zero. The other is by inserting
the eigenvalues using commutation methods. This works fine in case of
a constant background solution and the eigenvalues can be chosen arbitrarily.
However, in case of a (quasi-)periodic background solution
it turns out that the eigenvalues need to satisfy certain restrictions.
This was probably first observed in \cite{kumi}, where it was proven that
adding one eigenvalue to the two-gap Weierstrass solution of the
Korteweg-de Vries (KdV) equation preserves the asymptotics on one side,
but gives a phase shift on the other side. The general case was solved in
\cite{gesv}. In particular, this shows that the eigenvalues and
reflection coefficients can no longer be prescribed independently if
one wants to stay in the class of short-range perturbations of a given
quasi-periodic background. It turns out that these constraints are related
to the fact that the resolvent set of the background operator is not simply
connected in the (quasi-)periodic case. In this case we have to reconstruct
the transmission coefficient from its boundary values on this non simply
connected domain which is only possible in terms of multivalued functions
in general, see \cite{voza}. Hence one needs to impose algebraic constraints
on the scattering data to obtain a single-valued transmission coefficient.
It seems that this was first emphasized in \cite{emt}. 

The aim of the present paper is to make this reconstruction explicit
in terms of Abelian integrals on the underlying hyperelliptic Riemann
surface for the case of Jacobi operators (respectively the Toda equation).
However, similar results apply to one dimensional Schr\"odinger operators
(respectively the KdV equation). This will then allow us to derive an
explicit condition for single-valuedness and to establish trace formulas
which relate the scattering data to conserved quantities for the Toda hierarchy.
In particular, these trace formulas are extensions of well-known
sum rules (see e.g.\ \cite{caopt}, \cite{ks}, \cite{lns}, \cite{npy}, \cite{sizl}, \cite{zl})
which have attracted an enormous amount of interest recently.

To achieve this aim we will first compute the Green function, harmonic measure,
and Blaschke factors for our domain. This case seems
to be hard to find in the literature; the only example we could find is the
elliptic case in the book by Akhiezer \cite{ak}. See however also \cite{tom1},
\cite{tom2}, where similar questions are investigated.

\section{Notation}

To set the stage, let $\M$ be the Riemann surface associated with the function $\Rg{z}$,
where
\begin{equation}
R_{2g+2}(z) = \prod_{j=0}^{2g+1} (z-E_j), \qquad E_0 < E_1 < \cdots < E_{2g+1},
\end{equation}
$g\in \N$. $\M$ is a compact, hyperelliptic Riemann surface of genus $g$.
We will choose $\Rg{z}$ as the fixed branch
\begin{equation}
\Rg{z} = -\prod_{j=0}^{2g+1} \sqrt{z-E_j},
\end{equation}
where $\sqrt{.}$ is the standard root with branch cut along $(-\infty,0)$.

A point on $\M$ is denoted by $p = (z, \pm \Rg{z}) = (z, \pm)$, $z \in \C$.
The two points at infinity are denoted by $p = \infty_{\pm}$.
We use $\pi(p) = z$ for the projection onto the extended complex plane
$\C \cup \{\infty\}$.  The points $\{(E_{j}, 0), 0 \leq j \leq 2 g+1\} \subseteq \M$ are 
called branch points and the sets 
\begin{equation}
\Pi_{\pm} = \{ (z, \pm \Rg{z}) \mid z \in \C\backslash \Sigma\} \subset \M,
\qquad \Sigma= \bigcup_{j=0}^g[E_{2j}, E_{2j+1}],
\end{equation}
are called upper and lower sheet, respectively. Note that the boundary of
$\Pi_\pm$ consists of two copies of $\Sigma$ corresponding to the
two limits from the upper and lower half plane.

Let $\{a_j, b_j\}_{j=1}^g$ be loops on the Riemann surface $\M$ representing the
canonical generators of the fundamental group $\pi_1(\M)$. We require
$a_j$ to surround the points $E_{2j-1}$, $E_{2j}$ (thereby changing sheets
twice) and $b_j$ to surround $E_0$, $E_{2j-1}$ counter-clockwise on the
upper sheet, with pairwise intersection indices given by
\begin{equation}
a_j \circ a_k= b_j \circ b_k = 0, \qquad a_j \circ b_k = \delta_{jk},
\qquad 1 \leq j, k \leq g.
\end{equation}
The corresponding canonical basis $\{\zeta_j\}_{j=1}^g$ for the space of
holomorphic differentials can be constructed by
\begin{equation}
\underline{\zeta} = \sum_{j=1}^g \underline{c}(j)  
\frac{\pi^{j-1}d\pi}{R_{2g+2}^{1/2}},
\end{equation}
where the constants $\underline{c}(.)$ are given by
\[
c_j(k) = C_{jk}^{-1}, \qquad 
C_{jk} = \int_{a_k} \frac{\pi^{j-1}d\pi}{R_{2g+2}^{1/2}} =
2 \int_{E_{2k-1}}^{E_{2k}} \frac{z^{j-1}dz}{\Rg{z}} \in
\R.
\]
The differentials fulfill
\begin{equation}
\int_{a_j} \zeta_k = \delta_{j,k}, \qquad \int_{b_j} \zeta_k = \tau_{j,k}, 
\qquad \tau_{j,k} = \tau_{k, j}, \qquad 1 \leq j, k \leq g.
\end{equation}
For further information we refer to \cite{fk}, \cite[App.~A]{tjac}.

\section{Algebro-geometric constraints}

We are motivated by scattering theory for the pair $(H,H_q)$ of two Jacobi
operators, where $H$ is a short-range perturbation of a quasi-periodic
finite-gap operator $H_q$ associated with the Riemann surface introduced in
the previous section (see \cite[Ch.~9]{tjac}). One key quantity is the transmission
coefficient $T(z)$. It is meromorphic in $\Pi_+$ with finitely many simple poles
in $\Pi_+\cap\R$ precisely at the eigenvalues of the perturbed operator $H$. Since
\be \label{t2r2}
|T(\lam)|^2 + |R_\pm(\lam)|^2 =1, \qquad \lam\in \Sigma,
\ee
it can be reconstructed from the reflection coefficients $R_\pm(\lam)$
once we show how to reconstruct $T(z)$ from its boundary values
$|T(\lam)|^2=1-|R_\pm(\lam)|^2$, $\lam\in \partial\Pi_+$. Rather than enter
into more details here, see \cite{emt} (respectively \cite{voyu}),
we will focus on the reconstruction procedure only.

We begin by deriving a formula for the Green function of $\Pi_+$:

\begin{lemma}
The Green function of $\Pi_+$ with pole at $p_0$ is given by
\be
g(z,z_0) = - \re \int_{E_0}^p \om_{p_0 \ti{p}_0}, \quad p=(z,+),\: p_0=(z_0,+),
\ee
where $\ti{p}_0= \ol{p_0}^*$ (i.e., the complex conjugate on the other sheet)
and  $\om_{p q}$ is the normalized Abelian differential of the third kind with poles
at $p$ and $q$.
\end{lemma}

\begin{proof}
First of all observe $\om_{p_0 \ti{p}_0}= \om_{p_0 E_0} - \om_{\ti{p}_0 E_0}$ and
set
\be
\om_{p_0 E_0} = r_\pm(z,z_0) dz
\ee
on $\Pi_\pm$. Since $\om_{p_0 E_0}$ is continuous on the branch cuts,
the corresponding values of $r_\pm$ must match up, that is,
\be
\lim_{\eps\downarrow 0} r_+(\lam+\I\eps,z_0) =
\lim_{\eps\downarrow 0} r_-(\lam-\I\eps,z_0), \quad \lam\in\Sigma.
\ee
Moreover,
\be
\om_{\ti{p}_0 E_0} = \ol{r_\mp(\ol{z},z_0)} dz
\ee
on $\Pi_\pm$. Hence,
\be
\om_{p_0 \ti{p}_0} = \lim_{\eps\downarrow 0}
(r_+(\lam+\I\eps,z_0) - \ol{r_-(\lam-\I\eps,z_0)} d\lam =
2\I\,\im(r(\lam,z_0)) d\lam, \quad \lam \in\Sigma,
\ee
where $r(\lam,z_0)= \lim_{\eps\downarrow 0}r_+(\lam+\I\eps,z_0)$, shows
that $\om_{p_0,\ti{p}_0}$ is purely imaginary on the boundary of $\Pi_+$.
Together with the fact that the $a$-periods of $\om_{p_0 \ti{p}_0}$ vanish
this shows $\int_{E_0}^p \om_{p_0 \ti{p}_0}$ is purely imaginary on $\partial\Pi_+$.
Hence $g(z,z_0)$ vanishes on $\partial\Pi_+$ and since it has the proper singularity
at $z_0$ by construction, we are done.
\end{proof}

Clearly, we can extend $g(z,z_0)$ to a holomorphic function on $\M\backslash\{z_0\}$
by dropping the real part. By abuse of notation we will denote this function by
$g(p,p_0)$ as well. However, note that $g(p,p_0)$ will be multivalued with
jumps in the imaginary part across $b$-cycles. We will choose the path
of integration in $\C\backslash[E_0,E_{2g+1}]$ to guarantee a single-valued
function. 

From the Green's function we obtain the Blaschke factor and the harmonic
measure (see e.g., \cite{tsu}). Since we are mainly interested in the case
where the poles are on the real line (since $T(z)$ has all poles on the real line),
we note the following relation which will be needed later on:

\begin{lemma} \label{lemsygf}
For $\rho$ with $\pi(\rho)\in\R\backslash\Sigma$ we have
\be
g(p,\rho)= \int_{E_0}^p \om_{\rho \rho^*} = \int_{E(\rho)}^\rho \om_{p p^*},
\ee
where $E(\rho)$ is $E_0$ if $\rho<E_0$, either $E_{2j-1}$ or $E_{2j}$ if
$\rho\in(E_{2j-1},E_{2j})$, $1\le j \le g$, and $E_{2g+1}$ if $\rho>E_{2g+1}$.
\end{lemma}

\begin{proof}
By symmetry of the Green's function this holds at least when taking real parts.
Since both quantities are real for $\pi(p)<E_0$ it holds everywhere.
\end{proof}

Now we come to the Blaschke factor
\be
B(p,\rho)= \exp \Big( g(p,\rho) \Big) = \exp\Big(\int_{E_0}^p \om_{\rho \rho^*}\Big),
\qquad \pi(\rho)\in\R,
\ee
and first show that it can be written in terms of theta functions.

\begin{lemma}
The Blaschke factor is given by
\be
B(p,\rho)= \frac{\theta(\Amap(\rho^*) + \hamap(\di) - \vrc)}
{\theta(\Amap(\rho) + \amap(\di) - \vrc)}
\frac{\theta(\Amap(p) - \Amap(\rho) - \amap(\di) + \vrc)}{\theta(\Amap(p) - \Amap(\rho^*) - \hamap(\di) + \vrc)},
\ee
where $\di$ is any divisor of degree $g-1$ such that $\di_{\rho} + \di$ and
$\di_{\rho^*} + \di$ are nonspecial.

In addition, it satisfies
\be
B(E_0,\rho)=1, \quad\mbox{and}\quad
B(p^*,\rho) = B(p,\rho^*) = B(p,\rho)^{-1};
\ee
it is real-valued for $\pi(p)\in(-\infty,E_0)$.
\end{lemma}

\begin{proof}
Both the Blaschke factor and the quotient of theta functions are multivalued
meromorphic functions. Invoking the bilinear relations shows that
we have the same jumps $\int_{\ti{\rho}}^\rho \zeta_\ell$ around $b$-cycles.
Hence their quotient is single-valued. Moreover both have the same
divisor $\di_{\rho}-\di_{\ti{\rho}}$ and hence the quotient is holomorphic and
thus constant.

To see that $B(p^*,\rho)B(p,\rho)=1$, note that this function has no jumps and no poles
and hence is constant. Since it is one at $E_0$ it is one everywhere.
\end{proof}

Next we compute the harmonic measure of $\partial\Pi_+$.

\begin{lemma}
The harmonic measure of $\partial\Pi_+$ with pole at $\lam$ is given by
\be
\mu(p,\lam)d\lam = \frac{1}{\pi} \im\, \om_{p E_0}(\lam) =
-\frac{1}{\pi} \im\left( \int_{E_0}^p \om_{\lam,0} \right) d\lam,
\ee
where $\om_{\lam,0}$ is the normalized Abelian differential of the second kind with a
second-order pole at $\lam$.
\end{lemma}

\begin{proof}
All we have to do is to compute $(2\pi)^{-1} \lim_{\eps\downarrow 0}
\frac{\partial}{\partial\eps} g(z,\lam \pm \I\eps)$ (where the sign is chosen
according to which side of $\Sigma$ one is interested):
\be
- \frac{\partial}{\partial\eps} \re \int_{E_0}^{\lam\pm\I\eps} \om_{p \ti{p}} \Big|_{\eps=0}
= \im\, \om_{p \ti{p}} = 2 \im\, \om_{p E_0}
\ee
since $\om_{p \ti{p}} = \om_{p E_0} - \om_{\ti{p} E_0}$.

The other formula follows similarly:
\be
- \frac{\partial}{\partial\eps} \re \int_{E_0}^p \om_{p_0\ti{p}_0} \Big|_{\eps=0}
= -2 \im \int_{E_0}^p \om_{\lam,0}
\ee
since $\frac{\partial}{\partial z_0} \om_{p_0 E_0} = - \om_{z_0,0}$.
\end{proof}

Note that
$$
\om_{p E_0} = \left( \frac{-1}{2(\lam-E_0)} + o(1) \right) d\lam
$$
for $\lam$ near $E_0$ and that the imaginary part has no
singularity for $\lam\in\partial\Pi_+$.

Now we can characterize the scattering data (\cite{emt}, \cite{emt2}):

\begin{theorem}
Let $T(z)$ be meromorphic in $\Pi_+$ with simple poles at
$\{ \rho_j \}_{j=1}^q \subseteq \R\backslash\Sigma$ such that
$T$ is continuous up to the boundary with the only possible simple zeros
at the branch points.

Then $T(z)$ can be recovered from the boundary values $\ln|T(\lam)|$,
$\lam\in\Sigma$, via the Poisson-Jensen-type formula
\be \label{Tai}
T(z) = \left( \prod_{j=1}^q B(p,\rho_j)^{-1} \right)
\exp \left(\frac{1}{2\pi \I} \int_{\partial\Pi_+} 
\ln|T|^2 \om_{p E_0} \right), \quad p=(z,+),
\ee
where we have identified $\rho_j$ with $(\rho_j,+)$ and defined
$T(p)= \lim_{\eps\downarrow 0} T(\lam \pm\eps)$, $p=(\lam,\pm)\in\partial\Pi_+$.
\end{theorem}

\begin{proof}
The formula for $T(z)$ holds by \cite[Thm.~1]{voza}, when
taking absolute values. Since both sides are analytic and have equal
absolute values, they can only differ by a constant of absolute value one.
But both sides are positive at $z=\infty$ and hence this constant is one.
\end{proof}

\begin{remark}
A few remarks are in order:
\begin{enumerate}
\item
The integrand in (\ref{Tai}) is not integrable at $E_0$ and the integral
has to be understood as a principal value. Otherwise, one can move
the singularity away from $\partial\Pi_+$ which just alters the value by
a constant.
\item
In scattering theory one has $|T(p^*)|=|T(p)|$, $p\in\partial\Pi_+$,
and under this assumption we have
$$
T(z) = \left( \prod_{j=1}^q \exp\left(-\int_{E(\rho_j)}^{\rho_j} \om_{p p^*}\right) \right)
\exp \left(\frac{1}{2\pi \I} \int_\Sigma 
\ln(1 - |R_\pm|^2) \om_{p p^*} \right),
$$
where $E(\rho)$ is defined in Lemma~\ref{lemsygf} and the integral over
$\Sigma$ is taken on the upper sheet.
\item
The Abelian differential is explicitly given by
$$
\om_{p q} = \left( \frac{R_{2g+2}^{1/2} + \Rg{p}}{2(\pi - \pi(p))} -
\frac{R_{2g+2}^{1/2} + \Rg{q}}{2(\pi - \pi(q))} + P_{p q}(\pi) \right)
\frac{d\pi}{R_{2g+2}^{1/2}},
$$
where $P_{p q}(z)$ is a polynomial of degree $g-1$ which has to be determined from
the normalization $\int_{a_\ell} \om_{p p^*}=0$. In particular,
$$
\om_{p p^*} = \left( \frac{\Rg{p}}{\pi - \pi(p)} + P_{p p^*}(\pi) \right)
\frac{d\pi}{R_{2g+2}^{1/2}}.
$$
\end{enumerate}
\end{remark}

In inverse scattering theory one uses (\ref{t2r2}) to reconstruct $T$ from the
reflection coefficient $R_+$ (or $R_-$). Since neither the Blaschke factors
nor the outer function in (\ref{Tai}) are in general single-valued on $\Pi_+$,
we are naturally interested in when $T$ is single-valued for given $R_\pm$.

\begin{theorem}
The transmission coefficient $T$ defined via (\ref{Tai}) is single-valued if and
only if the eigenvalues $\rho_j$ and the reflection coefficient $R_\pm$
satisfy
\be \label{algcon}
\sum_j \int_{\rho_j^*}^{\rho_j} \zeta_\ell -
\frac{1}{2\pi\I} \int_{\partial \Pi_+}  \!\!\! \ln(1 - |R_{\pm}|^2) \zeta_\ell \in \Z.
\ee
\end{theorem}

\begin{proof}
For $T(z)$ to be single-valued we need $\lim_{\eps\downarrow 0} T(x-\I\eps)
= \lim_{\eps\downarrow 0} T(x+\I\eps)$ for every $x$ in a spectral gap.
If $x\in (E_{2\ell-1},E_{2\ell})$ is in the $\ell$'th gap, the path
of integration from $E_0$ to $\lam$ from above and back from $\lam$ to $E_0$
from below just gives the $b$-cycle $b_\ell$. Hence,
\be
\lim_{\eps\downarrow 0} \frac{T(x+\I\eps)}{T(x-\I\eps)} =
\exp \left(
\sum_{j=1}^q  \int_{b_\ell} \om_{\rho_j,\ti\rho_j} -
\frac{1}{2\pi\I} \int_{\partial\Pi_+} \!\!\!
\ln(1 - |R_{\pm}(\lam)|^2) \int_{b_\ell}\om_{\lam,0} d\lam \right).
\ee
Evaluating the $b_\ell$-cycle using the usual bilinear relations
finally yields
\be
\lim_{\eps\downarrow 0} \frac{T(x+\I\eps)}{T(x-\I\eps)} =
\exp \left(
2\pi\I \sum_{j=1}^q  \int_{\ti\rho_j}^{\rho_j} \zeta_\ell -
\int_{\partial\Pi_+}  \!\!\!
\ln(1 - |R_{\pm}|^2) \zeta_\ell \right)
\ee
and if the limit is supposed to be one, we are lead to (\ref{algcon}).
\end{proof}

\noindent
The special case for an elliptic background with zero reflection coefficient
was first obtained in \cite{kumi}. An analogous result was obtained in a different
context by \cite{tom1}. One should also emphasize that (\ref{algcon})
is only a necessary condition for the solvability of the inverse scattering problem.
Necessary and sufficient conditions are given in \cite{emt}.

\section{Trace formulas}

The transmission coefficient also plays a central role in the inverse
scattering transform. Since it turns out to be the perturbation determinant of
the pair $(H, H_q)$, in the sense of Krein (\cite{gokr}, \cite{krein}), its asymptotic
expansion provides the conserved quantities of the Toda hierarchy
(\cite{emt2}, \cite{tist}, \cite{mt}),
\be
\frac{d}{dz} \ln T(z) = - \sum_{k=1}^\infty \frac{\tau_k}{z^{k+1}}, \quad
\tau_k = \tr(H^k - (H_q)^k).
\ee
Relating this expansion with the one obtained by expanding (\ref{Tai}) near $z=\infty$,
one obtains the usual trace formulas (also known as Case-type sum
rules, \cite{caopt}).

Next, let $\om_j$ be the meromorphic differential
\be
\om_0 = \om_{\infty_+ \infty_-}, \quad
\om_k = \om_{\infty_+;k-1} - \om_{\infty_-;k-1},
\ee
where $\om_{p,k}$ is the Abelian differential of the second kind with a
pole of order $k+2$ at $p$. Note that $\om_k$ is of the form
\be
\om_k = \frac{P_k(\pi)}{R_{2g+2}^{1/2}} d\pi,
\ee
where $P_k(z)$ is a monic polynomial of degree $g+k$ whose coefficients have to
be determined from the fact that the $a$-cycles vanish and from the behavior at
$\infty_\pm$ (see \cite[Eq.~(13.30)]{tjac}).

\begin{theorem}
The following trace formulas are valid:
\bea\nn
\ln(T(\infty)) &=& -\sum_{j=1}^q \int_{E(\rho_j)}^{\rho_j} \om_{\infty_+ \infty_-}
+ \frac{1}{\pi\I} \int_\Sigma \ln|T|\, \om_{\infty_+ \infty_-},\\
\frac{1}{k+1} \tau_{k+1} &=& -\sum_{j=1}^q \int_{E(\rho_j)}^{\rho_j} \om_k
+ \frac{1}{\pi\I} \int_\Sigma \ln|T|\, \om_k,
\eea
where $E(\rho)$ is defined in Lemma~\ref{lemsygf} and the integral over
$\Sigma$ is taken on the upper sheet.
\end{theorem}

\begin{proof}
By $\frac{d^k}{dz^k} \omega_{p(z) E_0} |_{z=0} = k!\, \omega_{p_0,k-1}$, where
$z$ is a coordinate centered at $p_0$, we have
\be
\om_{p E_0} = \om_{\infty_+ E_0} + \sum_{k=1}^\infty z^k \omega_{p,k-1},
\qquad p=(\frac{1}{z},+),
\ee
and the claim follows.
\end{proof}

\bigskip

\noindent{\bf Acknowledgments.} I want to thank Peter Yuditskii for
several helpful discussions and for pointing out \cite{tom1,tom2,voza} to me.
In addition, I am also grateful to Fritz Gesztesy for discussions on this
topic.

\end{document}